\documentclass[prx,aps,reprint,showpacs,floatfix]{revtex4-1}
\usepackage{amsmath}
\usepackage{amsfonts}
\usepackage{amssymb}
\usepackage{revsymb}
\usepackage{graphicx}

\def\vec#1{\mathbf{#1}}

\def\eval#1{\left\langle #1 \right\rangle}

\begin{document}
\title{Prospects for atomic clocks based on large ion crystals.}
\author{Kyle Arnold}, 
\affiliation{Center for Quantum Technologies, 3 Science Drive 2, Singapore, 117543}
\affiliation{Department of Physics, National University of Singapore, 2 Science Drive 3, Singapore, 117551}
\author{Elnur Haciyev}
\affiliation{Center for Quantum Technologies, 3 Science Drive 2, Singapore, 117543}
\affiliation{Department of Physics, National University of Singapore, 2 Science Drive 3, Singapore, 117551}
\author{Eduardo Paez}
\affiliation{Center for Quantum Technologies, 3 Science Drive 2, Singapore, 117543}
\affiliation{Department of Physics, National University of Singapore, 2 Science Drive 3, Singapore, 117551}
\author{Chern Hui Lee}
\affiliation{Center for Quantum Technologies, 3 Science Drive 2, Singapore, 117543}
\affiliation{Department of Physics, National University of Singapore, 2 Science Drive 3, Singapore, 117551}
\author{John Bollinger}
\affiliation{Time and Frequency Division, National Institute of Standards and Technology, Boulder, CO 80305} 
\author{M. D. Barrett}
\email{phybmd@nus.edu.sg}
\affiliation{Center for Quantum Technologies, 3 Science Drive 2, Singapore, 117543}
\affiliation{Department of Physics, National University of Singapore, 2 Science Drive 3, Singapore, 117551}
\begin{abstract}
We investigate the feasibility of precision frequency metrology with large ion crystals.  For clock candidates with a negative differential static polarisability, we show that micromotion effects should not impede the performance of the clock.  Using Lu$^+$ as a specific example, we show that quadrupole shifts due to the electric fields from neighbouring ions do not significantly affect clock performance.  We also show that effects from the tensor polarisability can be effectively managed with a compensation laser at least for a small number of ions ($\lesssim 10^3$).  These results provide new possibilities for ion-based atomic clocks, allowing them to achieve stability levels comparable to neutral atoms in optical lattices and a viable path to greater levels of accuracy.
\end{abstract}
\maketitle
\section{Introduction}
The realisation of accurate, stable frequency references have enabled important advances in science and technology.  Well-known examples include the Global Positioning System, and tests of fundamental physical theories.  Increasing levels of accuracy and stability continue to be made with atomic clocks based on optical transitions in isolated atoms \cite{AlIon,SrYe,HgIon,SrIon,YbIon,InIon,YbForbidden,RMP,Ludlow}.  By now a number of groups have demonstrated superior performance over the current caesium frequency standards with the best clocks to date having inaccuracy at the $10^{-18}$ level \cite{AlIon,SrYe}.  For the past decade single ion clocks have held a leading position.  However in recent years, advances in laser stability have allowed neutral atoms to take advantage of large numbers of atoms giving superior performance in stability while maintaining some of the best accuracies \cite{SrYe2}.  Ion-based clocks have been limited to single atoms predominately due to the fact that trap influences such as micromotion are difficult to control for multiple ions.  Indeed, for the Al$^+$ clock, micromotion is a dominant factor in the overall error budget \cite{AlIon}.  We note that consideration has been given to clocks based on small strings of less than 10 ions stored in linear radio-frequency (RF) traps \cite{Mehlstauber}.

Micromotion is caused by the RF drive used to confine the ion.  It causes two correlated effects: the rapid oscillatory motion at the RF frequency $\Omega$ gives a second-order Doppler shift, and the electric field driving the motion induces an AC stark shift proportional to $\Delta\alpha$ where $\Delta \alpha=\alpha_e-\alpha_g$ is the differential static scalar polarisability.  In 1998 \cite{micromotion} it was noted that, when $\Delta \alpha<0$, these two effects could be made to exactly cancel for a well chosen value of $\Omega$. This value, which we refer to here as the magic RF frequency in analogy with the magic wavelength for optical lattices, depends only on the properties of the atom.   The existence of such a magic RF frequency gives rise to the important question of what happens in a large ion crystal where micromotion effects can be very pronounced at the edges of the crystal.  If the cancelation is maintained, then precision frequency metrology with large ion crystals is potentially feasible.

In this paper we show that dominant higher order micromotion effects can also be mitigated, and that clock operation would not be affected by micromotion.  For atoms with an electronic angular momentum $J>1/2$, we show that quadrupole shifts due to the electric fields from neighbouring ions do not significantly affect clock performance.  We also show that shifts arising from the tensor polarisability can be effectively compensated with an additional laser field.  Together these results show that ions having $\Delta \alpha<0$ can reap the benefits of large numbers of ions, just as neutral atoms can in optical lattices.  We illustrate our analysis using $^{176}$Lu$^+$ as a concrete example \cite{MDB1}, but the ideas can be readily adapted to any other ion with $\Delta\alpha<0$.

\section{Micromotion}
\label{Micromotion}
We start by writing the RF and static electric field potentials in the form
\begin{equation}
\phi_\mathrm{rf}=\frac{m\Omega \omega_{z}}{2q}\mathbf{r}^T \Lambda_\mathrm{rf} \mathbf{r}\cos{\Omega t},\quad \phi_\mathrm{s}=\frac{m\omega_z^2}{2q}\mathbf{r}^T \Lambda_\mathrm{s} \mathbf{r}
\end{equation}
where $\omega_z$ is one of the pseudo-potential oscillation frequencies, $\Omega$ is the RF drive frequency, and $m$ and $q$ are the mass and charge of the ion respectively.  The matrices $\Lambda_\mathrm{rf}$ and $\Lambda_s$ determine the curvatures of the potentials and in general, we may choose $\Lambda_\mathrm{rf}$ to be diagonal. Defined in this way, the pseudo-potential approximation is
\begin{equation}
V(\mathbf{r})=\frac{1}{2}m\omega_z^2 \mathbf{r}^T\left(\Lambda_\mathrm{s}+\frac{1}{2}\Lambda_\mathrm{rf}^2\right)\mathbf{r}
\end{equation}
If we scale time by $2/\Omega$ and length by 
\[
l=\left(\frac{q^2}{4\pi\epsilon_0 m \omega_z^2}\right)^{1/3},
\]
then the equations of motion (e.o.m.) are given by
\begin{equation}
\ddot{\vec{r}}_i+\left(\epsilon^2\Lambda_\mathrm{s}+2\epsilon\Lambda_\mathrm{rf}\cos2t\right)\vec{r}_i-\epsilon^2\sum_{j\neq i} \frac{\vec{r}_{ij}}{r_{ij}^3}=0
\end{equation}
where $\epsilon=2\omega_z/\Omega$.
Following the treatment in \cite{CrystalsI} we assume a stable $\pi$-periodic crystal solution exists, which may be expressed as a Fourier expansion
\begin{equation}
\vec{r}_i^\pi(t)=\sum_{n=-\infty}^{n=\infty} \vec{R}_{2n,i} e^{2 i n t}.
\end{equation}
In this form, $\vec{R}_{0,i}$ is the time-averaged position of the $i^\mathrm{th}$ ion.  Substituting this expansion into the e.o.m. and using a Taylor series expansion for the Coulomb term about $\vec{R}_{0,ij}=\vec{R}_{0,i}-\vec{R}_{0,j}$ we can obtain an infinite set of coupled equations for $\vec{R}_{2n,i}$ representing the Fourier expansion of the e.o.m.  Defining
\begin{equation}
\label{def1}
\vec{F}_{ij}=\frac{\vec{R}_{0,ij}}{|\vec{R}_{0,ij}|^3}, \quad Q_{ij}=-\frac{3\vec{R}_{0,ij}\vec{R}_{0,ij}^T-|\vec{R}_{0,ij}|^2}{|\vec{R}_{0,ij}|^5},
\end{equation}
the first two Fourier equations are given by
\begin{equation}
\label{n0eq}
\epsilon^2\Lambda_\mathrm{s} \vec{R}_{0,i}+2\epsilon\Lambda_\mathrm{rf}\vec{R}_{2,i}-\epsilon^2\sum_{j\neq i} \vec{F}_{ij}=0
\end{equation}
and 
\begin{multline}
\label{n1eq}
\left(\epsilon^2\Lambda_\mathrm{s}-4\mathbb{I}\right) \vec{R}_{2,i}+\epsilon\Lambda_\mathrm{rf}\left(\vec{R}_{0,i}+\vec{R}_{4,i}\right)\\
-\epsilon^2\sum_{j\neq i} Q_{ij}\left(\vec{R}_{2,i}-\vec{R}_{2,j}\right)=0.
\end{multline}
To lowest order, Eq.~\ref{n1eq} gives
\begin{equation}
\label{amp1}
\vec{R}_{2,i}=\frac{\epsilon}{4}\Lambda_\mathrm{rf}\vec{R}_{0,i}
\end{equation}
which expresses the fact that the micro-motion amplitude is directly proportional to the RF electric field at the position of the ion.  With this approximation, the fractional shift of a clock transition is given by
\begin{equation}
\label{mm0}
\frac{\Delta\nu_i}{\nu}=-\left(\frac{\omega_z l}{2 c}\right)^2\left[1+\frac{\Delta \alpha}{h \nu}\left(\frac{m\Omega c}{q}\right)^2\right]\vec{R}_{0,i}^T \Lambda_\mathrm{rf}^2\vec{R}_{0,i}.
\end{equation}
For $\Delta \alpha<0$ this leads to a magic RF drive frequency defined by
\begin{equation}
\label{magic0}
\Omega_0=\frac{q}{mc}\sqrt{\frac{h \nu}{-\Delta \alpha}}
\end{equation}
at which micro-motion shifts cancel as first pointed out in \cite{micromotion}.  For one ion, Eq.~\ref{mm0} is sufficient for even the very best clocks \cite{AlIon,HgIon,SrIon}.  However, for large ion crystals, higher order corrections should be considered.  To this purpose, we first note that the $n=2$ Fourier component of the e.o.m. is
\begin{equation}
\label{harmonic}
\vec{R}_{4,i}=\frac{\epsilon}{16}\Lambda_\mathrm{rf}\vec{R}_{2,i},
\end{equation}
to lowest order.  Substitution into Eq.~\ref{n1eq} then gives
\begin{multline}
\label{n1eqApprox}
\left(\epsilon^2\left(\Lambda_\mathrm{s}+\frac{1}{16}\Lambda_\mathrm{rf}^2\right)-4\mathbb{I}\right) \vec{R}_{2,i}+\epsilon\Lambda_\mathrm{rf}\vec{R}_{0,i}\\
-\epsilon^2\sum_{j\neq i} Q_{ij}\left(\vec{R}_{2,i}-\vec{R}_{2,j}\right)=0.
\end{multline}
Using the fact that $\left(\mathbb{I}-\epsilon^2 A\right)^{-1}\approx \mathbb{I}+\epsilon^2 A$, we can solve for $\vec{R}_{2,i}$ to get
\begin{multline}
\label{amp2}
\vec{R}_{2,i}=\frac{1}{4}\left(\mathbb{I}+\frac{\epsilon^2}{4}\left(\Lambda_\mathrm{s}+\frac{1}{16}\Lambda_\mathrm{rf}^2\right)\right)\epsilon \Lambda_\mathrm{rf}\vec{R}_{0,i}\\
-\frac{\epsilon^3}{16}\sum_{j\neq i} Q_{ij} \Lambda_\mathrm{rf}\left(\vec{R}_{0,i}-\vec{R}_{0,j}\right).
\end{multline}
The term on the second line of Eq.~\ref{amp2} is the coupling of the micromotion amplitudes of each ion through the Coulomb interaction.  Physically it arises from a distortion of the space-charge potential due to differential micromotion amplitudes between ions, which provides an effective RF electric field in addition to the trap drive.  This effective electric field also provides an additional AC stark shift.

The electric field at the $i^\mathrm{th}$ ion due to the space charge is
\begin{equation}
\label{spaceQ}
\mathbf{E}_i = \frac{m \omega_z^2 l}{q}\sum_{j\neq i} \left(\mathbf{F}_{ij} + 2 Q_{ij} (\vec{R}_{2,i}-\vec{R}_{2,j})\cos(2 t)\right).
\end{equation}
Using Eq.~\ref{amp1} and defining
\begin{equation}
\label{Wdef}
\vec{W}_{0,i}=\sum_{j\neq i}Q_{ij} \Lambda_\mathrm{rf} (\vec{R}_{0,i}-\vec{R}_{0,j}),
\end{equation}
the amplitude of the net RF electric field on the $i^\mathrm{th}$ ion is then
\begin{equation}
\label{totRF}
\mathbf{E}_{i,\mathrm{RF}} = -\frac{m \omega_z\Omega l}{q} \left(\Lambda_\mathrm{rf}\vec{R}_{0,i} - \frac{\epsilon^2}{4}\vec{W}_{0,i}\right).
\end{equation}

Using Eq.~\ref{amp2}, Eq.~\ref{totRF} and noting that $\left(\mathbb{I}+\epsilon^2 A\right)^2 \approx \mathbb{I}+2\epsilon^2 A$ the fractional frequency shift of a clock transition due to terms oscillating at the RF drive frequency is then
\begin{multline}
\label{mm1}
\frac{\Delta\nu_i}{\nu}=-\left(\frac{\omega_z l}{2 c}\right)^2\Bigg\{\left[1-\left(\frac{\Omega}{\Omega_0}\right)^2\right]\vec{R}_{0,i}^T \Lambda_\mathrm{rf}^2\vec{R}_{0,i}\\
-\frac{\epsilon^2}{2}\left[1-\left(\frac{\Omega}{\Omega_0}\right)^2\right]\vec{R}_{0,i}^T \Lambda_\mathrm{rf}\vec{W}_{0,i}\\
+\frac{\epsilon^2}{2}\vec{R}_{0,i}^T\Lambda_\mathrm{rf}\left( \Lambda_\mathrm{s}+\frac{1}{16}\Lambda_\mathrm{rf}^2\right)\Lambda_\mathrm{rf}\vec{R}_{0,i}\Bigg\}.
\end{multline}

To the same order of approximation, we must include a DC stark shift from the space charge and a time dilation shift from the higher harmonic $\vec{R}_{4,i}$.  These can be added independently, which follows from the orthogonality of the Fourier components.  At equilibrium, the sum of all DC fields on the $i^\mathrm{th}$ ion exactly balances the pseudo-potential force from the RF field so we have
\begin{equation}
\vec{E}_{i,\mathrm{DC}}=\frac{1}{2}\frac{m \omega_z^2 l}{q} \Lambda_\mathrm{rf}^2 \vec{R}_{0,i}.
\end{equation}
This gives a fractional frequency shift
\begin{eqnarray}
\frac{\Delta\nu_i}{\nu}&=&-\frac{\Delta \alpha}{2h\nu}\left(\frac{m \omega_z^2 l}{2q}\right)^2\mathbf{R}_{0,i}^T \Lambda_\mathrm{rf}^4 \mathbf{R}_{0,i}\nonumber\\
\label{spaceDC}
&=&\frac{\epsilon^2}{8}\left(\frac{\omega_z l}{2c}\right)^2\left(\frac{\Omega}{\Omega_0}\right)^2\mathbf{R}_{0,i}^T \Lambda_\mathrm{rf}^4 \mathbf{R}_{0,i}\
\end{eqnarray}
and the time dilation shift from $\vec{R}_{4,i}$ is
\begin{equation}
\label{harmonic}
\frac{\Delta\nu_i}{\nu}=-\frac{\epsilon^2}{64}\left(\frac{\omega_z l}{2c}\right)^2\mathbf{R}_{0,i}^T \Lambda_\mathrm{rf}^4 \mathbf{R}_{0,i}.
\end{equation}
The shifts given in Eq.~\ref{spaceDC} \&~\ref{harmonic} can then be added to Eq.~\ref{mm1} to give the total shift correct to $\mathcal{O}(\epsilon^2)$.  

In addition to a fractional frequency shift, micromotion also gives rise to an effective frequency modulation of the probe; a fact that can be used to detect micromotion \cite{micromotion}.  The effect reduces the probe coupling to $J_0(\beta)$ where $J_0$ is the first order Bessel function and the modulation index is given by $\beta=2l\mathbf{k}\cdot \mathbf{R}_{2,i}$. Because the micromotion amplitude can be very large for ions far removed from the zero point of the RF field \cite{micromotion}, the ability of the probe to drive the clock transition can be significantly diminished.  This can be avoided by probing along the RF null axis of a linear Paul trap, for which there is very little micromotion.

For the linear Paul trap, we can write
\begin{equation}
\Lambda_\mathrm{rf}=\begin{pmatrix} a & 0 & 0\\0 & -a & 0\\0 & 0 & 0\end{pmatrix},\; \Lambda_\mathrm{s}=\begin{pmatrix} -\frac{1}{2}+\delta & 0 & 0\\0 & -\frac{1}{2}-\delta & 0\\0 & 0 & 1\end{pmatrix},
\end{equation}
where $a$ determines the strength of the RF confinement relative to the DC field.  The parameter $\delta$ determines the asymmetry in the transverse dimension and can be tuned arbitrarily by appropriate choice of biasing voltages.  Writing $\Lambda_\mathrm{rf}=a\Lambda$, the total shift from Eq.~\ref{mm1},~\ref{spaceDC}, \&~\ref{harmonic} can be written
\begin{multline}
\label{mm2}
\frac{\Delta\nu_i}{\nu}=-\left(\frac{a \omega_z l}{2 c}\right)^2\Bigg\{\left[\lambda_0-\left(\frac{\Omega}{\Omega_0}\right)^2\right]\lambda_1\vec{R}_{0,i}^T \Lambda^2\vec{R}_{0,i}\\
\mspace{-3mu}+\frac{\delta \epsilon^2}{2}\vec{R}_{0,i}^T\Lambda\vec{R}_{0,i}-\frac{\epsilon^2}{2 a}\left[1-\left(\frac{\Omega}{\Omega_0}\right)^2\right]\vec{R}_{0,i}^T \Lambda\vec{W}_{0,i}\Bigg\},
\end{multline}
where the $\lambda_k$ are given by
\begin{equation}
\lambda_0=1-\frac{16+5 a^2}{32}\frac{\epsilon^2}{2 \lambda_1},\mbox{ and } \lambda_1=1+\frac{a^2\epsilon^2}{8}.
\end{equation}
The first term of Eq.~\ref{mm2} is zero for $\Omega=\Omega_0\sqrt{\lambda_0}$ and this can be viewed as a correction to the magic RF frequency given by Eq.~\ref{magic0}.  It also applies to the single ion case even though this term includes the shift due to the DC component of the space charge.  This is because it is the space charge that provides the static electric field necessary to displace an ion to the equilibrium position $\vec{R}_{0,i}$.  The second term also applies to the single ion case and is only present when the transverse confinement is non-degenerate.  The third term is only applicable to the many ion case as it is a consequence of the induced RF field from the oscillating space charge.

When probing a clock transition, micromotion will give rise to an inhomogeneous broadening of the line and a shift of the line centre. Since the first term in Eq.~\ref{mm2} can be tuned to zero, the inhomogeneous broadening is limited only by the remaining terms and the degree of broadening is determined by the size of the crystal.  The second term is suppressed for small $\delta$ and, at $\Omega=\Omega_0\sqrt{\lambda_0}$, the final term scales as $\epsilon^4$.  If we can neglect the inhomogeneous broadening, the shift of the line centre is then determined by the average of Eq.~\ref{mm2}.  For large crystals we can assume the density of ions is approximately constant and the average of Eq.~\ref{mm2} can be taken as an integral over a continuum.  The shift of the clock transition then scales as $N^{2/3}$, where $N$ is the number ions and the scale factor depends on the geometry of the trap \cite{Haase}.  This leads to a further modification to the magic RF drive frequency at which the scale factor vanishes.  In a practical application we would simply vary the number of ions and tune the variation of the clock frequency to zero by adjustment of $\Omega$.

For reasons to be discussed in the next section, it is advantageous to take a spherically symmetric trap for which $a=\sqrt{3}$ and $\delta=0$.  In the spherically symmetric case, an analytic approximation for $\vec{W}_{0,i}$ can be obtained. By taking the continuum limit with a constant density of ions, we can approximate the sum in Eq.~\ref{Wdef} by an integral and we obtain
\begin{equation}
\label{RFapprox}
\vec{W}_{0,i}\approx -\frac{1}{5} \Lambda_\mathrm{rf} \vec{R}_{0,i}.
\end{equation}
In this case the fractional frequency shift takes the simple form
\begin{equation}
\label{mm3}
\frac{\Delta\nu_i}{\nu}=-\left(\frac{a \omega_z l}{2 c}\right)^2\left[\lambda_0^\prime-\left(\frac{\Omega}{\Omega_0}\right)^2\right]\lambda_1^\prime\vec{R}_{0,i}^T \Lambda^2\vec{R}_{0,i}
\end{equation}
where the $\lambda_k^\prime$ are given by
\begin{equation}
\lambda_0^\prime=1-\frac{31}{64}\frac{\epsilon^2}{\lambda_1^\prime},\mbox{ and } \lambda_1^\prime=1+\frac{19}{40} \epsilon^2.
\end{equation}
Note that $\lambda_0-\lambda_0^\prime\sim \mathcal{O}(\epsilon^4)$ indicating that the oscillating space charge has no significant affect to the accuracy of the treatment given.

At this point it is useful to illustrate our analysis with an example.  We simulate the distribution of ions by integrating the e.o.m.\ in the pseudo-potential approximation including a small damping term to anneal the initial state.  For the initial state we use multi-layered Mackay icosahedra \cite{Mackay}  with a random offset to each particle coordinate of about $10\%$ of the initial minimum particle spacing.  With $N=100$ to $5000$ ions we were able to confirm the results given in \cite{Haase}.  In particular, to a good approximation we have
\begin{equation}
\eval{\vec{R}_{0,i}^T \Lambda^2\vec{R}_{0,i}}=\frac{2}{5}N^{2/3}-0.3964.
\end{equation} 

\begin{figure}
\begin{center}
\includegraphics[width=8cm]{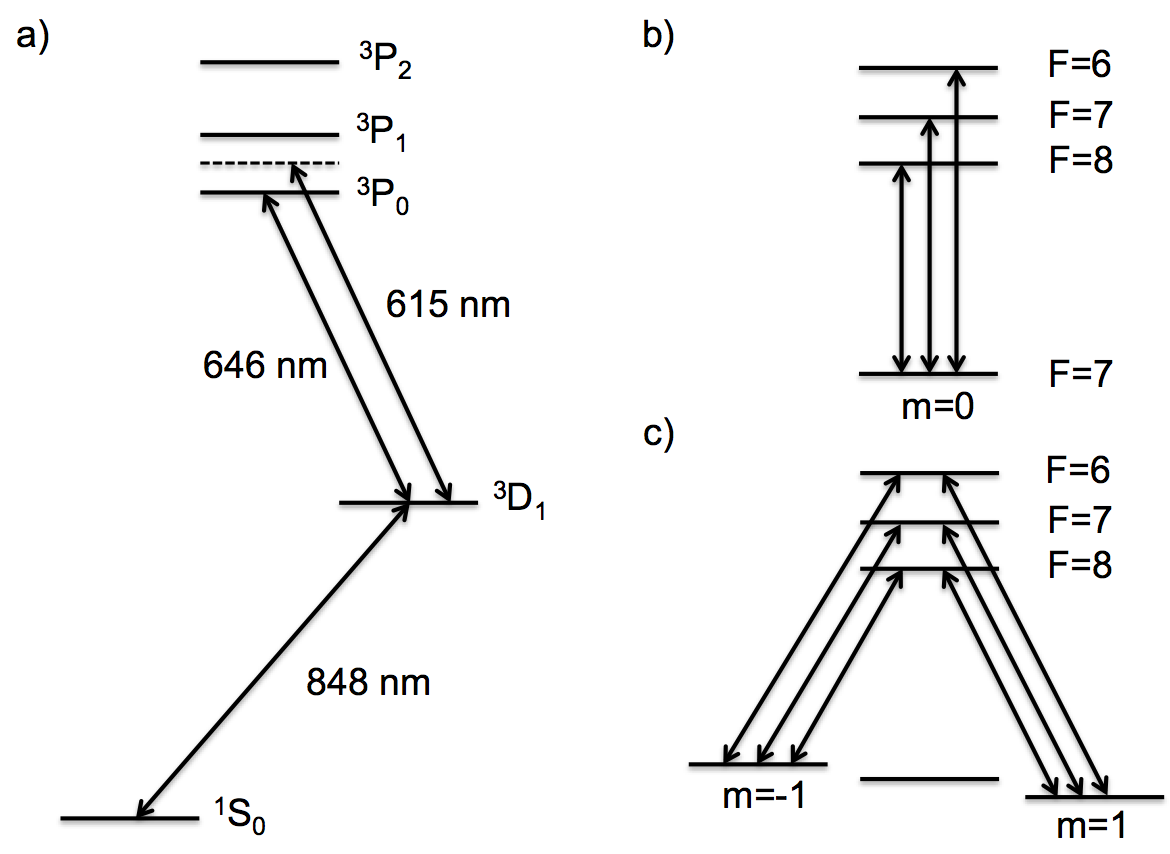}
\caption{(a) Level structure of $^{176}$Lu$^+$, which has a nuclear spin $I=7$, showing the clock transition at $848\,\mathrm{nm}$ and the cooling and detection transition at $646\,\mathrm{nm}$.  A laser at $615\,\mathrm{nm}$ provides a magic wavelength to compensate tensor polarisability shifts from the RF field. (b) Transitions need for hyperfine averaging. (c) Transitions needed to effectively realise the transitions in (b) due to constraints on the quantisation axis - see Sec.~\ref{Tensor}}
\label{LuStructure}
\end{center}
\end{figure}

For the ion properties we use $^{176}$Lu$^+$ which has a level structure as illustrated in Fig.~\ref{LuStructure}.  For $^{176}$Lu$^+$, the estimated value of the differential static scalar polarisability is $\Delta\alpha=-2.19$ a.u. \cite{Dzuba}\footnote{Throughout we use atomic units for polarisabilities.  Conversion to S.I. units is via the scale factor $4\pi \epsilon_0 a_0^3$ where $a_0$ is the Bohr radius}, giving a magic RF frequency of $\Omega_0\approx2\pi\times 23.2\,\mathrm{MHz}$ \cite{Calc1}.   We take $\omega_z=2\pi\times 200\,\mathrm{kHz}$ which, for $^{176}$Lu$^+$, gives $l=7.94\,\mathrm{\mu m}$.
So, for the parameters given, we have
\begin{equation}
\left(\frac{a \omega_z l}{2 c}\right)^2\approx 8.3\times 10^{-16},\quad \epsilon^2 \approx 3.0\times10^{-4}.
\end{equation}
With $\Omega=\Omega_0$ and $N=5000$ ions,  the higher order terms result in a broadening of $3.3\times10^{-17}$ and an average shift of $1.4\times 10^{-17}$.  These values would scale as $N^{2/3}$.  When $\Omega=\Omega_0\sqrt{\lambda_0^\prime}$  the average shift vanishes and this is only a $7\times10^{-5}$ fractional change to the RF drive frequency.  Comparison of the clock frequency at different numbers would therefore provide an accurate assessment of $\Omega$ and $\Delta\alpha$.

From the above analysis it would appear micromotion would not limit the accuracy that can be achieved.  Moreover, the small broadening effects would not affect achievable stability in the foreseeable future.  As the crystal size increases, so does the demands on the pointing stability of the probe but, for $10^4$ ions and an angular misalignment of $0.1^\circ$, the Rabi frequency for the outermost ions would be diminished by just $3\times 10^{-4}$.  

Our analysis has only considered the scalar polarisability.  Levels with $J=1/2$ have a vector polarisability but this is only relevant for circular polarisations.  For candidates such as Lu$^+$, which have clock states with $J>1/2$, we must also consider effects arising from the quadrupole moment and the tensor polarisability.
\section{Considerations for $\mathbf{J>1/2}$}
In \cite{MDB1}, Barrett showed that averaging over transitions to all hyperfine states of a fixed $m_F\leq I-J$ cancels dominant magnetic field effects and quadrupole shifts whenever the nuclear spin, $I$, is at least as large as $J$.  This averaging, which we shall refer to as hyperfine averaging, is very general and applies to any perturbation that can be described by a rank $k>0$ tensor operator that does not depend on $I$.  Hence, we need only consider the inhomogeneous broadening arising from such interactions.  The following considerations can also be readily adapted for candidates with $I=0$, such as $^{88}$Sr$^+$ for example \cite{SrIon}, which utilise averaging over all $m_J$.  
\subsection{Quadrupole Shifts}
\label{Quadrupole}
Since the quadrupole shift from the DC trapping field is fixed, it does not contribute to any broadening and so we only consider the quadrupole fields arising from the space charge.  Since the quadrupole field from neighbouring ions falls off cubicly with distance, the quadrupole field experienced by an ion is due mostly to its local environment.  Except for ions near to the edge of the crystal we can anticipate that the local environment is essentially the same for each ion and the resulting quadruple shift is relatively homogeneous.  Indeed for sufficiently large numbers, it is known that a crystal of long range order forms \cite{Haase,Drewsen,Bollinger1} and the limiting structure is that of a body-centered cubic (bcc) lattice.  In this regime, the quadrupole shift would be constant for the bulk of the crystal and by symmetry it would be zero.

The tensor describing the quadrupole field for the $i^\mathrm{th}$ ion due to all other ions is given by
\begin{equation}
 \mathbf{\nabla E}^{(2)}_i =\frac{m \omega_z^2}{e} Q_i= \frac{m \omega_z^2}{e}\sum_{j\neq i} Q_{ij},
\end{equation}
where $Q_{ij}$ is given in Eq.~\ref{def1}.  Itano \cite{ItanoQuad} has derived the quadrupole shift for a general orientation of the quantisation axis relative to the coordinate system for which $Q_i$ is given.  This shift factors into a geometric term, a state dependent scale factor $C_{F,m_F}$ on the order of unity, and an overall scale factor quantifying the size of the shift.  The geometrical factor is given by
\begin{multline}
\label{quadfac}
\frac{Q_{zz}}{4}\left(3\cos^2 \beta-1\right)+\frac{1}{2} \sin (2\beta) \left(Q_{xz}\cos\alpha+Q_{yz}\sin\alpha\right)\\
+\frac{1}{4} \sin^2 \beta  \big((Q_{xx}-Q_{yy})\cos(2\alpha)+2Q_{xy}\sin(2\alpha)\big),
\end{multline}
where we have used the Euler angle definitions in \cite{ItanoQuad} and dropped the subscript $i$ for convenience. For the $m_F=0$ states of the $^3\mathrm{D}_1$ level of $^{176}\mathrm{Lu}^+$, the state dependent scale factors are $-2/5, 1, -3/5$ for the $F=6,7,8$ hyperfine levels, respectively.  In the calculations that follow we omit this factor.  The overall scale factor depends on the quadrupole moment.  For the estimated value of $-1.3 e a_0^2$ \cite{Dzuba}, its size is $-1.3 m\omega_z^2 a_0^2 \approx h\times 2.5\,\mathrm{Hz}$ where $a_0$ is the Bohr radius.

\begin{figure}
\begin{center}
\includegraphics{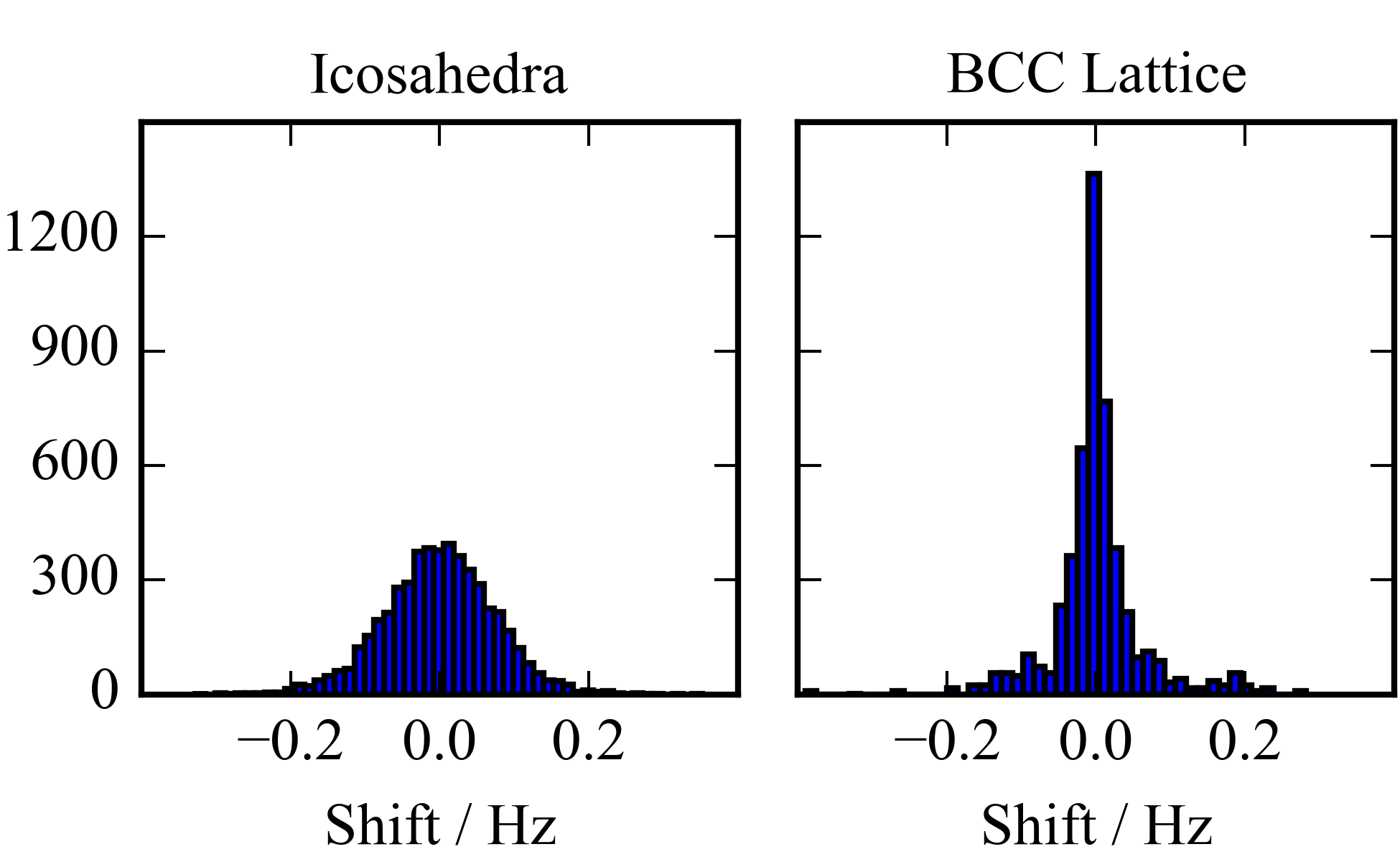}
\caption{Distribution of quadrupole shifts for 5000 ions in a spherically symmetric trap.  Initial starting distributions are a multi-layered Mackay Icosahedra (left) and body-centered cubic lattice (right).}
\label{Fig1}
\end{center}
\end{figure}

In Fig.~\ref{Fig1} we plot the distribution of quadrupole shifts for $N=5000$ for a spherically symmetric trap.  As in \cite{Haase}, we do see a dependence of the final crystal configuration on the initial condition for larger $N$.  Configurations starting from a bcc-lattice tend to stay in this configuration with some rounding near the boundary of the crystal.  Hence we give the distribution of quadrupole shifts for two types of initial conditions: a multi-layered Mackay icosahedra applicable to smaller numbers of ions, and a bcc lattice applicable in the limit of large $N$.  The distribution on the left is more applicable to smaller numbers as considered here and is approximately Gaussian with a standard deviation of $0.078\,\mathrm{Hz}$.  The distribution does not depend on $N$ and we have verified that there is no significant dependence on the field orientation as expected from the spherical symmetry. This level of the broadening should not be an issue even for the very best lasers available today.

The distribution on the right is applicable to larger numbers of ions. However crystals of long range order have been reported for smaller numbers \cite{Drewsen}.  Comparing the two distributions we clearly see the effect of the bcc-lattice component giving the expected peaking of the distribution near zero. This distribution does have a dependence on the orientation of the B-field and we have oriented the field along one of the axes of the interior bcc structure which coincides with the trap axis \footnote{The alignment of the crystal here is by construction of the initial condition.}.  We would expect the dependence on the orientation of the B-field to diminish for larger numbers as the bcc component becomes much more prominent.

The distribution of quadrupole shifts depends only on the geometry of the crystal and not on its overall size, at least for the range of numbers we have explored.  Prolate ellipsoidal crystals in more conventional linear Paul traps, in which the transverse confinement is much stronger than the axial confinement, have a much broader distribution of quadrupole shifts.  Moreover, the width of the distribution depends on the orientation of the trap relative to the quantisation axis as may be expected.  For these reasons, we have restricted our attention to a spherical geometry.
\subsection{Tensor Polarisability}
\label{Tensor}
The tensor polarisability also gives rise to a shift of the clock frequency from the RF fields.  As shown in \cite{ItanoQuad,polarisability}, this contribution is given by
\begin{equation}
\label{tensor}
\frac{\delta \nu}{\nu}=-\frac{C_{F,m_F}}{4}\frac{\alpha_{2,J}}{h \nu} \langle3 E_z^2-E^2\rangle,
\end{equation}
where $C_{F,m_F}$ is a state dependent scale factor identical to those for the quadrupole shift, $\langle\cdot\rangle$ indicates a time average over one cycle of the oscillating field, and the tensor polarisability $\alpha_{2,J}$ is in general frequency dependent.  Since the RF frequency is small relative to any optical frequency of interest, we can use the DC value of $\alpha_{2,J}\approx -5.0$ a.u. for the tensor polarisability \cite{Dzuba}.  If the quantisation axis is aligned along the trap axis, then $E_z=0$ and the shift has the same form as discussed for the scalar polarisability.  In general, due to the $F$ dependent scale factors, we cannot simply modify the scalar polarisability to account for the effect \footnote{For a single transition, if $\alpha_{2,J}$ was sufficiently small, it could be incorporated into the definition of $\Delta \alpha$}.  Since Eq.~\ref{tensor} applies at all frequencies, we can use a laser field to reduce the broadening that arises.  We also note that the constraint on the quantisation axis forces us to realise $m_F=0$ to $m_F=0$ transitions as an average over $m_F=\pm 1$ to  $m_F=0$ as illustrated in Fig.~\ref{LuStructure}.  This introduces further averaging to that discussed in \cite{MDB1} but does not affect any of the points considered here.

Use of a laser to reduce the broadening arising from the tensor polarisability requires the spatial dependence of the beam to match the spatial variation of the RF field and the use of a magic wavelength at which the dynamic differential scalar polarisability of the clock transition is zero.  For Lu$^+$, such a wavelength can be found for a laser tuned between the $^3\mathrm{D}_1$ to $^3\mathrm{P}_0$ and $^3\mathrm{D}_1$ to $^3\mathrm{P}_1$ transitions.  From matrix elements given in \cite{lifetimes2}, we find a magic wavelength at $\approx 615\,\mathrm{nm}$ with $\alpha_{2,J}\approx100$ a.u.  The sign of $\alpha_{2,J}$ relative to the DC value requires the use of a doughnut mode \cite{doughnut} which, to lowest order, has an intensity profile that has the same quadratic dependence on the distance from the trap axis as the $E^2$ amplitude of the RF field.  In this case, off-resonant scattering from the compensation beam is completely determined by the amount of compensation needed.  For a crystal of $10^4$ ions, we estimate a scattering rate $\approx 0.005\,\mathrm{s}^{-1}$ for the outermost ions.  Hence, for Lu$^+$, off-resonant scattering will not limit this approach.  A more practical limit would be the mode matching of the laser profile to the spatial variation of the RF field.  But we emphasise that the mode matching need only be sufficient to reduce the broadening since hyperfine averaging eliminates any residual shift.  

\begin{figure}
\begin{center}
\includegraphics{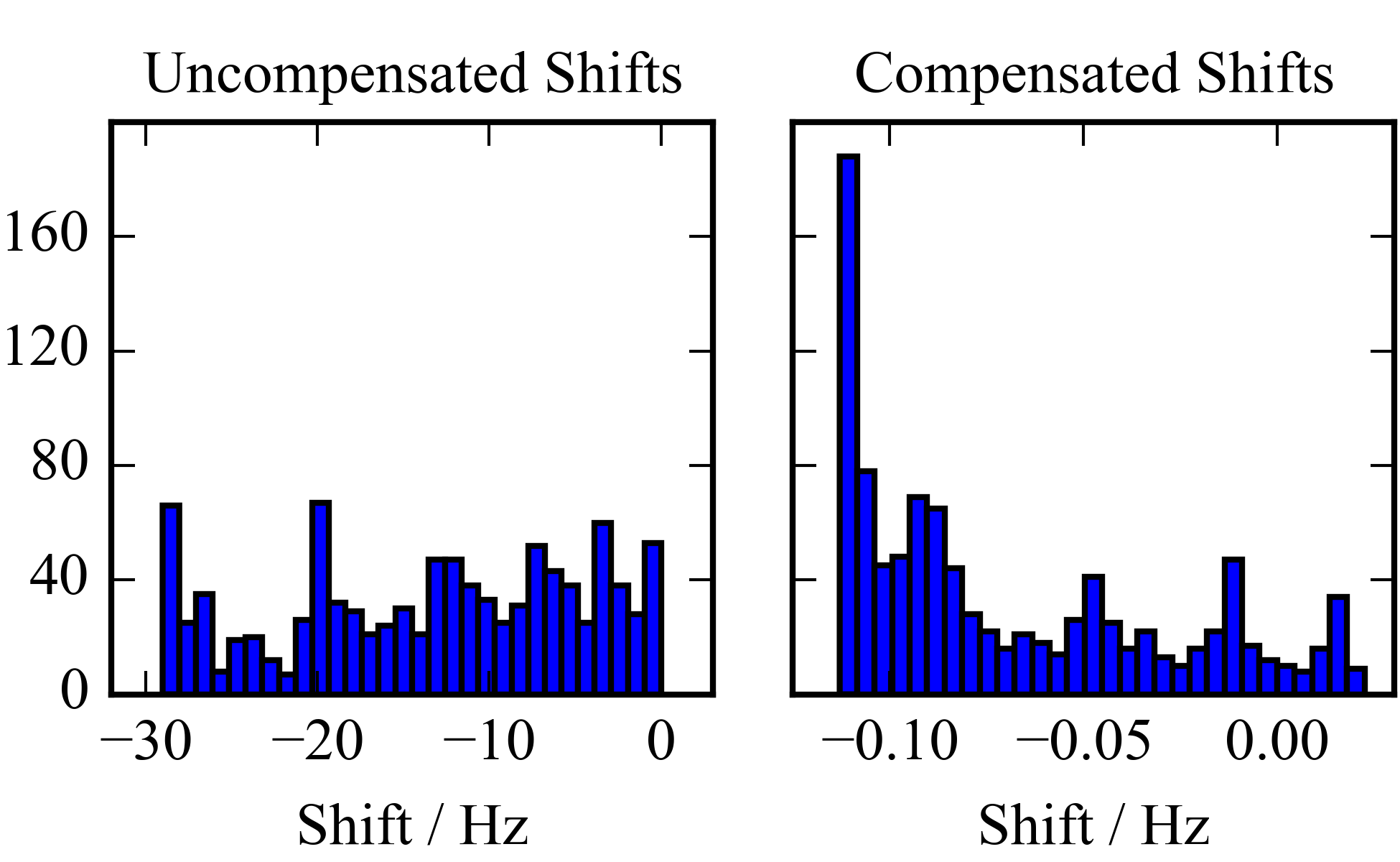}
\vspace{-0.25cm}
\caption{Distribution of shifts due to the tensor polarisability for $N=1000$ ions.  Distribution on the right has been compensated using a doughnut Laguerre-Gauss beam with a waist of $100l\approx 800\,\mathrm{\mu m}$.  We have used the same spherical geometry as in the previous sections.}
\vspace{-0.5cm}
\label{tensorShifts}
\end{center}
\end{figure}

In Fig.~\ref{tensorShifts} we illustrate the compensation of broadening for $N=1000$ ions using a doughnut Laguerre-Gauss beam with a waist of $100l\approx 800\,\mathrm{\mu m}$ with the same spherical trap geometry used in the previous sections.  Due to the fact that the width of the broadening scales as $N^{2/3}$ it would likely become impractical to go beyond $N=1000$ ions. This assessment is based on due consideration of demonstrated mode purity of higher order Laguerre-Gauss beams \cite{doughnut}, and realistic constraints on beam size.
\section{Prospects for Lu$^{+}$}
As a clock candidate, Lu$^+$ has a number of favourable properties leading to low systematic shifts which are summarised in Table~\ref{table1}.  The blackbody radiation shift at $300\,\mathrm{K}$ is based on the current estimate of the differential static polarisability of $\Delta\alpha=-2.19\,\mathrm{a.u.}$ \cite{Dzuba}.  The second-order Doppler shift due to residual secular thermal motion assumes Doppler limited cooling on the $646\,\mathrm{nm}$ transition.  This has a line-width of $\Gamma=2\pi\times 2.45 \,\mathrm{MHz}$ providing one of the lowest Doppler cooling limits amongst the ions and yet sufficiently large to allow a collection of $>5$ photons/ms per ion during detection.  Hyperfine averaging cancels dominate Zeeman shifts leaving only a residual quadratic shift of $\approx 5\,\mathrm{Hz/mT^2}$ due to coupling to the $^3$D$_2$ fine structure level \cite{MDB1}.  The shift given in the table assumes an operating field of $10\,\mathrm{\mu T}$.  Finally, AC Stark shifts from the probe laser are based on current estimates of the dynamic polarisabilities at the clock frequency ($\Delta \alpha_0=-18.12$ a.u. and $\alpha_{2}=-12.44$ a.u.), and a lifetime of $62\,\mathrm{h}$ \cite{Dzuba}.  The inequality given in the table applies to all six transitions involved in the hyperfine averaging as indicated in Fig.~\ref{LuStructure}(c).
\begin{table}
\begin{ruledtabular}
\centering
\begin{tabular}{l c}
Effect & Shift $(10^{-18})$\vspace{2pt}\\
\hline
Blackbody Radiation @ 300K & 53.3\\
Secular Doppler & -0.05\\
Micromotion & 0\\
Quadrupole Shifts & 0\\
Quadratic Zeeman @ $10\,\mathrm{\mu T}$ & -1.4\\
Probe AC Stark (200 ms $\pi$-pulse) & $< 50$\\
\end{tabular}
\caption{Summary of systematic shifts for Lu$^+$ after hyperfine averaging.}
\label{table1}
\end{ruledtabular}
\end{table}

We can expect much of the AC Stark shift from the clock laser to be eliminated by hyper-Ramsey spectroscopy \cite{hyperRamsey1, hyperRamsey2}.  Thus a temperature inaccuracy of $1$ degree at room temperature would permit fractional inaccuracies below $10^{-18}$.  Within a cryogenic environment we can also anticipate inaccuracies beyond $10^{-19}$.

As discussed in Sec.~\ref{Tensor}, we can expect to be limited in practice to $N \approx 1000$ ions.  For this many ions, if we combine the quadrupole shifts and the compensated shifts from the tensor polarisability, we obtain a reasonably symmetric distribution which can be roughly approximated to a Gaussian with a standard deviation of about $0.1\,\mathrm{Hz}$.  With the actual distribution, a simulated Ramsey experiment with a $T_m=1\,\mathrm{s}$ free precession time yields an 80\% contrast in the Ramsey fringes.  For Lu$^+$ state preparation and detection can be expected to take $\sim 1\%$ of the total interrogation time and hence the Dick effect \cite{Dick} should not have a significant role.  Neglecting the slight loss in fringe contrast, gives an estimated projection noise limited stability \cite{RMP, Itano} of
\begin{equation}
\label{stability}
\sigma(\tau)=\frac{1}{2\pi \nu_0\sqrt{N T_m \tau}}\approx\frac{1.5\times 10^{-17}}{\sqrt{\tau}}.
\end{equation}

With a stability given by Eq.~\ref{stability}, measurement of the clock frequency at levels of $10^{-18}$ could be achieved within $\approx 5$ minutes.  Assessment of micromotion shifts associated with small inaccuracies of the magic RF frequency does, however, require comparison of clock measurements with different numbers of ions.  Integration times of approximately 1 day would permit assessment of the clock at the $2\times 10^{-19}$ level for 100 ions.  Subsequent comparison with $1000$ ions would then provide a measurement accuracy of $\Delta \alpha$ and the magic RF frequency at the $10^{-6}$  level.  Using micromotion shifts to determine $\Delta \alpha$ has been demonstrated with a single ion \cite{MagicSr}.  With many ions, the sensitivity of this approach is substantially improved.

\section{Alternative traps}
Our approach has focused on linear Paul traps.  Other approaches may be feasible such as multipole traps or Penning traps.  In multipole traps low numbers of ions initially populate a single 1-dimensional ring of ions that can hold up to several tens of ions.  By symmetry quadrupole shifts and shifts from the tensor polarisability are practically constant for all ions.  As more ions are added, more rings form. For more than 2 rings, the shifts split into multiple values that can be separated at the Hertz level.  Thus it would be difficult to go beyond a few $10^2$ ions by this approach.  Nevertheless, this could be achieved with very little broadening thus allowing for much longer Ramsey times.

Penning traps have been used to confine and control very large numbers of ions \cite{Bollinger1}.  Ion confinement in a Penning trap is due to the ion crystal rotation through a large uniform magnetic field.  In a Penning trap the combined fractional frequency shift is
\begin{equation}
\label{mmpt}
\frac{\Delta\nu}{\nu}=-\frac{1}{2}\left(\frac{\omega_r}{c}\right)^2\left[1+\frac{\Delta \alpha}{h \nu}\left(\frac{m\omega_r c}{e}\right)^2\right]\rho^2,
\end{equation}
where $\rho$ is the cylindrical radius of an ion in the crystal and $\omega_r$ is the rotation frequency of the crystal.  This then leads to a magic rotation frequency analogous to the magic RF frequency for the RF Paul trap.  For the Penning trap there are no higher-order corrections so the 2nd order Doppler and polarization compensation should work very well, but there are constraints on $\omega_r$ due to available magnetic fields.  Specifically, the rotation frequency is bounded by the cyclotron frequency $\Omega_c=e B/m$ \cite{Bollinger2}. This leads to the constraint
\begin{equation}
B>\sqrt{\frac{h \nu}{-c^2\Delta \alpha}}.
\end{equation}
Unfortunately this favours a large polarisability and small clock frequency.  Even a clock frequency as low as $10^{14}\,\mathrm{Hz}$ and a differential polarisability $\Delta \alpha=-100$ requires $B>22\,\mathrm{T}$.
\section{Conclusion}
We have shown that ions with a negative differential static polarisability should allow high precision metrology on large ion crystals.  More specifically, we have shown that micromotion does not give rise to any significant inhomogeneous broadening and that higher order frequency shifts can be managed through adjustment of the magic RF frequency.  For clock candidates that support a quadrupole moment, we have shown that spherically symmetric traps show very little broadening due to quadrupole shifts induced by neighbouring ions and this broadening is not dependent on the size of the crystal.  We have also shown that broadening arising from the tensor polarisability can be compensated by a laser field, at least for smaller numbers of ions ($\lesssim 1000$).  This will extend the advantage of using large numbers to ion-candidates having $\Delta\alpha<0$; an advantage that has allowed neutral atoms to surpass the performance of single ion standards.  In the case of $^{176}$Lu$^+$, this approach could outperform the current state of the art by an order of magnitude in both stability and accuracy.

We have not included effects due to anharmonicities of the trapping fields as these effects are design dependent.  However the framework we have used follows that given in \cite{CrystalsI} and should allow such effects to be included given a particular design.  The effect will be to give a spatial dependence to $\Lambda_\mathrm{s}$ and $\Lambda_\mathrm{rf}$.  But the influence of $\Lambda_\mathrm{s}$ only appears to second order in $\epsilon$ and we have shown that these effects do not contribute significantly under most circumstances.  Furthermore, the spatial dependence of $\Lambda_\mathrm{rf}$ in Eq.~\ref{amp1} would not change the lowest-order equation for the magic RF frequency.  Hence, we believe that the main effect of anharmonicity will be to affect the higher order terms only.  This would introduce a small amount of broadening and not change the general conclusions we have made here.  We may anticipate a variation in crystal density giving further broadening due to quadrupole shifts, but this would likely only be significant for highly anharmonic confinement.

We have also not considered magnetic field inhomogeneities as these are again design dependent. Magnetic fields arising from currents induced by the trap driving field \cite{AlIon} can be expected to have a significant spatial variation giving rise to a broadening through the quadratic Zeeman shift of the clock states.  Static field inhomogeneities would also give rise to additional broadening through the linear Zeeman effect.   Thus candidates with small B-field sensitivities would be desirable, but this is true of any clock.  Moreover, it may be possible in a specific implementation to compensate any significant broadening with additional fields as we have shown for tensor polarisability effects. 

Although the importance of $\Delta\alpha<0$ was pointed out over fifteen years ago \cite{micromotion}, it has not played a significant role in the development of ion-based atomic clocks.  This is perhaps due to the scarcity of candidates having this property.  To our knowledge there are eight candidates that have been reported in the literature: B$^+$ \cite{Safranova} , Ca$^+$ \cite{Safranova}, Sr$^+$ \cite{Safranova}, Ba$^+$ \cite{Sahoo}, Ra$^+$ \cite{Sahoo}, Er$^{2+}$\cite{Calc1}, Tm$^{3+}$\cite{Calc1}, and Lu$^+$\cite{Calc1}.  Of these candidates, B$^+$ is the only candidate with a $J=0$ to $J=0$ clock transition for which quadrupole and tensor polarisability restrictions do not apply.  However, the magic RF frequency for B$^+$ is $\approx 800\,\mathrm{MHz}$ which may be technically challenging to implement.  Moreover, the only cooling and detection channel available is the $^1$S$_0$ to $^1$P$_1$ at $137\,\mathrm{nm}$.  

From the alkaline-earth metals, Ba$^+$ is an interesting possibility.  For $^{137}$Ba$^+$, there are a number of states of the D$_{3/2}$ level with $C_{F,m_F}=0$.  Hence, quadrupole and tensor polarisability considerations would not apply.  Since these are the main limitations to working with large numbers, very high levels of stability could be possible with this ion. It is hoped that this discussion spurs interest in finding new candidate transitions with $\Delta\alpha<0$.

\begin{acknowledgements}
We would like to thank Vladimir Dzuba, for providing us with the latest matrix element calculations for Lu$^+$, and Alex Kozlov for useful discussions.  We would like to thank David Hume and Kyle Beloy for careful reading of the manuscript and their thoughtful comments.  We acknowledge the support of this work by the National Research Foundation and the Ministry of Education of Singapore. This manuscript is a contribution of NIST and is not subject to US copyright.
\end{acknowledgements}
\bibliographystyle{apsrev4-1}
\bibliography{ManyIonClock}

\begin{thebibliography}{35}%
\makeatletter
\providecommand \@ifxundefined [1]{%
 \@ifx{#1\undefined}
}%
\providecommand \@ifnum [1]{%
 \ifnum #1\expandafter \@firstoftwo
 \else \expandafter \@secondoftwo
 \fi
}%
\providecommand \@ifx [1]{%
 \ifx #1\expandafter \@firstoftwo
 \else \expandafter \@secondoftwo
 \fi
}%
\providecommand \natexlab [1]{#1}%
\providecommand \enquote  [1]{``#1''}%
\providecommand \bibnamefont  [1]{#1}%
\providecommand \bibfnamefont [1]{#1}%
\providecommand \citenamefont [1]{#1}%
\providecommand \href@noop [0]{\@secondoftwo}%
\providecommand \href [0]{\begingroup \@sanitize@url \@href}%
\providecommand \@href[1]{\@@startlink{#1}\@@href}%
\providecommand \@@href[1]{\endgroup#1\@@endlink}%
\providecommand \@sanitize@url [0]{\catcode `\\12\catcode `\$12\catcode
  `\&12\catcode `\#12\catcode `\^12\catcode `\_12\catcode `\%12\relax}%
\providecommand \@@startlink[1]{}%
\providecommand \@@endlink[0]{}%
\providecommand \url  [0]{\begingroup\@sanitize@url \@url }%
\providecommand \@url [1]{\endgroup\@href {#1}{\urlprefix }}%
\providecommand \urlprefix  [0]{URL }%
\providecommand \Eprint [0]{\href }%
\providecommand \doibase [0]{http://dx.doi.org/}%
\providecommand \selectlanguage [0]{\@gobble}%
\providecommand \bibinfo  [0]{\@secondoftwo}%
\providecommand \bibfield  [0]{\@secondoftwo}%
\providecommand \translation [1]{[#1]}%
\providecommand \BibitemOpen [0]{}%
\providecommand \bibitemStop [0]{}%
\providecommand \bibitemNoStop [0]{.\EOS\space}%
\providecommand \EOS [0]{\spacefactor3000\relax}%
\providecommand \BibitemShut  [1]{\csname bibitem#1\endcsname}%
\let\auto@bib@innerbib\@empty
\bibitem [{\citenamefont {Chou}\ \emph {et~al.}(2010)\citenamefont {Chou},
  \citenamefont {Hume}, \citenamefont {Koelemeij}, \citenamefont {Wineland},\
  and\ \citenamefont {Rosenband}}]{AlIon}%
  \BibitemOpen
  \bibfield  {author} {\bibinfo {author} {\bibfnamefont {C.~W.}\ \bibnamefont
  {Chou}}, \bibinfo {author} {\bibfnamefont {D.~B.}\ \bibnamefont {Hume}},
  \bibinfo {author} {\bibfnamefont {J.~C.~J.}\ \bibnamefont {Koelemeij}},
  \bibinfo {author} {\bibfnamefont {D.~J.}\ \bibnamefont {Wineland}}, \ and\
  \bibinfo {author} {\bibfnamefont {T.}~\bibnamefont {Rosenband}},\ }\href@noop
  {} {\bibfield  {journal} {\bibinfo  {journal} {Phys. Rev. Lett.}\ }\textbf
  {\bibinfo {volume} {104}},\ \bibinfo {pages} {070802} (\bibinfo {year}
  {2010})}\BibitemShut {NoStop}%
\bibitem [{\citenamefont {Bloom}\ \emph {et~al.}(2014)\citenamefont {Bloom},
  \citenamefont {Nicholson}, \citenamefont {Williams}, \citenamefont
  {Campbell}, \citenamefont {Bishof}, \citenamefont {Zhang}, \citenamefont
  {Zhang}, \citenamefont {Bromley},\ and\ \citenamefont {Ye}}]{SrYe}%
  \BibitemOpen
  \bibfield  {author} {\bibinfo {author} {\bibfnamefont {B.~J.}\ \bibnamefont
  {Bloom}}, \bibinfo {author} {\bibfnamefont {T.~L.}\ \bibnamefont
  {Nicholson}}, \bibinfo {author} {\bibfnamefont {J.~R.}\ \bibnamefont
  {Williams}}, \bibinfo {author} {\bibfnamefont {S.~L.}\ \bibnamefont
  {Campbell}}, \bibinfo {author} {\bibfnamefont {M.}~\bibnamefont {Bishof}},
  \bibinfo {author} {\bibfnamefont {X.}~\bibnamefont {Zhang}}, \bibinfo
  {author} {\bibfnamefont {W.}~\bibnamefont {Zhang}}, \bibinfo {author}
  {\bibfnamefont {S.~L.}\ \bibnamefont {Bromley}}, \ and\ \bibinfo {author}
  {\bibfnamefont {J.}~\bibnamefont {Ye}},\ }\href@noop {} {\bibfield  {journal}
  {\bibinfo  {journal} {Nature.}\ }\textbf {\bibinfo {volume} {506}},\ \bibinfo
  {pages} {71} (\bibinfo {year} {2014})}\BibitemShut {NoStop}%
\bibitem [{\citenamefont {Stalnaker}\ \emph {et~al.}(2007)\citenamefont
  {Stalnaker}, \citenamefont {Diddams}, \citenamefont {Fortier}, \citenamefont
  {Kim}, \citenamefont {Hollberg}, \citenamefont {Bergquist}, \citenamefont
  {Itano}, \citenamefont {Delany}, \citenamefont {Lorini}, \citenamefont
  {Oskay}, \citenamefont {Heavner}, \citenamefont {Jefferts}, \citenamefont
  {Levi}, \citenamefont {Parker},\ and\ \citenamefont {Shirley}}]{HgIon}%
  \BibitemOpen
  \bibfield  {author} {\bibinfo {author} {\bibfnamefont {J.~E.}\ \bibnamefont
  {Stalnaker}}, \bibinfo {author} {\bibfnamefont {S.}~\bibnamefont {Diddams}},
  \bibinfo {author} {\bibfnamefont {T.}~\bibnamefont {Fortier}}, \bibinfo
  {author} {\bibfnamefont {K.}~\bibnamefont {Kim}}, \bibinfo {author}
  {\bibfnamefont {L.}~\bibnamefont {Hollberg}}, \bibinfo {author}
  {\bibfnamefont {J.}~\bibnamefont {Bergquist}}, \bibinfo {author}
  {\bibfnamefont {W.}~\bibnamefont {Itano}}, \bibinfo {author} {\bibfnamefont
  {M.}~\bibnamefont {Delany}}, \bibinfo {author} {\bibfnamefont
  {L.}~\bibnamefont {Lorini}}, \bibinfo {author} {\bibfnamefont
  {W.}~\bibnamefont {Oskay}}, \bibinfo {author} {\bibfnamefont
  {T.}~\bibnamefont {Heavner}}, \bibinfo {author} {\bibfnamefont
  {S.}~\bibnamefont {Jefferts}}, \bibinfo {author} {\bibfnamefont
  {F.}~\bibnamefont {Levi}}, \bibinfo {author} {\bibfnamefont {T.}~\bibnamefont
  {Parker}}, \ and\ \bibinfo {author} {\bibfnamefont {J.}~\bibnamefont
  {Shirley}},\ }\href@noop {} {\bibfield  {journal} {\bibinfo  {journal} {Appl.
  Phys. B}\ }\textbf {\bibinfo {volume} {89}},\ \bibinfo {pages} {167}
  (\bibinfo {year} {2007})}\BibitemShut {NoStop}%
\bibitem [{\citenamefont {Dube}\ \emph {et~al.}(2013)\citenamefont {Dube},
  \citenamefont {Madej}, \citenamefont {Zhou},\ and\ \citenamefont
  {Bernard}}]{SrIon}%
  \BibitemOpen
  \bibfield  {author} {\bibinfo {author} {\bibfnamefont {P.}~\bibnamefont
  {Dube}}, \bibinfo {author} {\bibfnamefont {A.~A.}\ \bibnamefont {Madej}},
  \bibinfo {author} {\bibfnamefont {Z.}~\bibnamefont {Zhou}}, \ and\ \bibinfo
  {author} {\bibfnamefont {J.~E.}\ \bibnamefont {Bernard}},\ }\href@noop {}
  {\bibfield  {journal} {\bibinfo  {journal} {Phys. Rev. A.}\ }\textbf
  {\bibinfo {volume} {87}},\ \bibinfo {pages} {023806} (\bibinfo {year}
  {2013})}\BibitemShut {NoStop}%
\bibitem [{\citenamefont {Huntemann}\ \emph
  {et~al.}(2012{\natexlab{a}})\citenamefont {Huntemann}, \citenamefont
  {Okhapkin}, \citenamefont {Lipphardt}, \citenamefont {Weyers}, \citenamefont
  {Tamm},\ and\ \citenamefont {Peik}}]{YbIon}%
  \BibitemOpen
  \bibfield  {author} {\bibinfo {author} {\bibfnamefont {N.}~\bibnamefont
  {Huntemann}}, \bibinfo {author} {\bibfnamefont {M.}~\bibnamefont {Okhapkin}},
  \bibinfo {author} {\bibfnamefont {B.}~\bibnamefont {Lipphardt}}, \bibinfo
  {author} {\bibfnamefont {S.}~\bibnamefont {Weyers}}, \bibinfo {author}
  {\bibfnamefont {C.}~\bibnamefont {Tamm}}, \ and\ \bibinfo {author}
  {\bibfnamefont {E.}~\bibnamefont {Peik}},\ }\href@noop {} {\bibfield
  {journal} {\bibinfo  {journal} {Phys. Rev. Lett.}\ }\textbf {\bibinfo
  {volume} {108}},\ \bibinfo {pages} {090801} (\bibinfo {year}
  {2012}{\natexlab{a}})}\BibitemShut {NoStop}%
\bibitem [{\citenamefont {Wang}\ \emph {et~al.}(2007)\citenamefont {Wang},
  \citenamefont {Dumke}, \citenamefont {Liu}, \citenamefont {Stejskal},
  \citenamefont {Zhao}, \citenamefont {Zhang}, \citenamefont {Lu},
  \citenamefont {Wang}, \citenamefont {Becker},\ and\ \citenamefont
  {Walther}}]{InIon}%
  \BibitemOpen
  \bibfield  {author} {\bibinfo {author} {\bibfnamefont {Y.~H.}\ \bibnamefont
  {Wang}}, \bibinfo {author} {\bibfnamefont {R.}~\bibnamefont {Dumke}},
  \bibinfo {author} {\bibfnamefont {T.}~\bibnamefont {Liu}}, \bibinfo {author}
  {\bibfnamefont {A.}~\bibnamefont {Stejskal}}, \bibinfo {author}
  {\bibfnamefont {Y.}~\bibnamefont {Zhao}}, \bibinfo {author} {\bibfnamefont
  {J.}~\bibnamefont {Zhang}}, \bibinfo {author} {\bibfnamefont
  {Z.}~\bibnamefont {Lu}}, \bibinfo {author} {\bibfnamefont {L.~J.}\
  \bibnamefont {Wang}}, \bibinfo {author} {\bibfnamefont {T.}~\bibnamefont
  {Becker}}, \ and\ \bibinfo {author} {\bibfnamefont {H.}~\bibnamefont
  {Walther}},\ }\href@noop {} {\bibfield  {journal} {\bibinfo  {journal} {Opt.
  Commun.}\ }\textbf {\bibinfo {volume} {273}},\ \bibinfo {pages} {526}
  (\bibinfo {year} {2007})}\BibitemShut {NoStop}%
\bibitem [{\citenamefont {Barber}\ \emph {et~al.}(2006)\citenamefont {Barber},
  \citenamefont {Hoyt}, \citenamefont {Oates}, \citenamefont {Hollberg},
  \citenamefont {Taichenachev},\ and\ \citenamefont {Yudin}}]{YbForbidden}%
  \BibitemOpen
  \bibfield  {author} {\bibinfo {author} {\bibfnamefont {Z.~W.}\ \bibnamefont
  {Barber}}, \bibinfo {author} {\bibfnamefont {C.~W.}\ \bibnamefont {Hoyt}},
  \bibinfo {author} {\bibfnamefont {C.~W.}\ \bibnamefont {Oates}}, \bibinfo
  {author} {\bibfnamefont {L.}~\bibnamefont {Hollberg}}, \bibinfo {author}
  {\bibfnamefont {A.~V.}\ \bibnamefont {Taichenachev}}, \ and\ \bibinfo
  {author} {\bibfnamefont {V.~I.}\ \bibnamefont {Yudin}},\ }\href@noop {}
  {\bibfield  {journal} {\bibinfo  {journal} {Phys. Rev. Lett.}\ }\textbf
  {\bibinfo {volume} {96}},\ \bibinfo {pages} {083002} (\bibinfo {year}
  {2006})}\BibitemShut {NoStop}%
\bibitem [{\citenamefont {Ludlow}\ \emph {et~al.}(2014)\citenamefont {Ludlow},
  \citenamefont {Boyd}, \citenamefont {Peik},\ and\ \citenamefont
  {Schmidt}}]{RMP}%
  \BibitemOpen
  \bibfield  {author} {\bibinfo {author} {\bibfnamefont {A.~D.}\ \bibnamefont
  {Ludlow}}, \bibinfo {author} {\bibfnamefont {M.~M.}\ \bibnamefont {Boyd}},
  \bibinfo {author} {\bibfnamefont {E.}~\bibnamefont {Peik}}, \ and\ \bibinfo
  {author} {\bibfnamefont {P.}~\bibnamefont {Schmidt}},\ }\href@noop {}
  {\enquote {\bibinfo {title} {Optical atomic clocks},}\ } (\bibinfo {year}
  {2014}),\ \Eprint {http://arxiv.org/abs/physics.atom-th/1407.3493}
  {arXiv:physics.atom-th/1407.3493} \BibitemShut {NoStop}%
\bibitem [{\citenamefont {Hinkley}\ \emph {et~al.}(2013)\citenamefont
  {Hinkley}, \citenamefont {Sherman}, \citenamefont {Phillips}, \citenamefont
  {Schioppo}, \citenamefont {Lemke}, \citenamefont {Beloy}, \citenamefont
  {Pizzocaro}, \citenamefont {Oates},\ and\ \citenamefont {Ludlow}}]{Ludlow}%
  \BibitemOpen
  \bibfield  {author} {\bibinfo {author} {\bibfnamefont {N.}~\bibnamefont
  {Hinkley}}, \bibinfo {author} {\bibfnamefont {J.~A.}\ \bibnamefont
  {Sherman}}, \bibinfo {author} {\bibfnamefont {N.~B.}\ \bibnamefont
  {Phillips}}, \bibinfo {author} {\bibfnamefont {M.}~\bibnamefont {Schioppo}},
  \bibinfo {author} {\bibfnamefont {N.~D.}\ \bibnamefont {Lemke}}, \bibinfo
  {author} {\bibfnamefont {K.}~\bibnamefont {Beloy}}, \bibinfo {author}
  {\bibfnamefont {M.}~\bibnamefont {Pizzocaro}}, \bibinfo {author}
  {\bibfnamefont {C.~W.}\ \bibnamefont {Oates}}, \ and\ \bibinfo {author}
  {\bibfnamefont {A.~D.}\ \bibnamefont {Ludlow}},\ }\href@noop {} {\bibfield
  {journal} {\bibinfo  {journal} {Science}\ }\textbf {\bibinfo {volume}
  {341}},\ \bibinfo {pages} {1215} (\bibinfo {year} {2013})}\BibitemShut
  {NoStop}%
\bibitem [{\citenamefont {Nicholson}\ \emph {et~al.}(2015)\citenamefont
  {Nicholson}, \citenamefont {Campbell}, \citenamefont {Hutson}, \citenamefont
  {Marti}, \citenamefont {Bloom}, \citenamefont {McNally}, \citenamefont
  {Zhang}, \citenamefont {Barrett}, \citenamefont {Safronova}, \citenamefont
  {Strouse},\ and\ \citenamefont {andJ. Ye}}]{SrYe2}%
  \BibitemOpen
  \bibfield  {author} {\bibinfo {author} {\bibfnamefont {T.~L.}\ \bibnamefont
  {Nicholson}}, \bibinfo {author} {\bibfnamefont {S.~L.}\ \bibnamefont
  {Campbell}}, \bibinfo {author} {\bibfnamefont {R.~B.}\ \bibnamefont
  {Hutson}}, \bibinfo {author} {\bibfnamefont {G.~E.}\ \bibnamefont {Marti}},
  \bibinfo {author} {\bibfnamefont {B.~J.}\ \bibnamefont {Bloom}}, \bibinfo
  {author} {\bibfnamefont {R.~L.}\ \bibnamefont {McNally}}, \bibinfo {author}
  {\bibfnamefont {W.}~\bibnamefont {Zhang}}, \bibinfo {author} {\bibfnamefont
  {M.~D.}\ \bibnamefont {Barrett}}, \bibinfo {author} {\bibfnamefont {M.~S.}\
  \bibnamefont {Safronova}}, \bibinfo {author} {\bibfnamefont {G.}~\bibnamefont
  {Strouse}}, \ and\ \bibinfo {author} {\bibfnamefont {W.~L.~T.}\ \bibnamefont
  {andJ. Ye}},\ }\href@noop {} {\bibfield  {journal} {\bibinfo  {journal}
  {Nature. Comm.}\ }\textbf {\bibinfo {volume} {6}},\ \bibinfo {pages} {6896}
  (\bibinfo {year} {2015})}\BibitemShut {NoStop}%
\bibitem [{\citenamefont {Herschbach}\ \emph {et~al.}(2012)\citenamefont
  {Herschbach}, \citenamefont {Pyka}, \citenamefont {Keller},\ and\
  \citenamefont {Mehlstauber}}]{Mehlstauber}%
  \BibitemOpen
  \bibfield  {author} {\bibinfo {author} {\bibfnamefont {N.}~\bibnamefont
  {Herschbach}}, \bibinfo {author} {\bibfnamefont {K.}~\bibnamefont {Pyka}},
  \bibinfo {author} {\bibfnamefont {J.}~\bibnamefont {Keller}}, \ and\ \bibinfo
  {author} {\bibfnamefont {T.~E.}\ \bibnamefont {Mehlstauber}},\ }\href@noop {}
  {\bibfield  {journal} {\bibinfo  {journal} {Appl. Phys. B.}\ }\textbf
  {\bibinfo {volume} {107}},\ \bibinfo {pages} {891} (\bibinfo {year}
  {2012})}\BibitemShut {NoStop}%
\bibitem [{\citenamefont {Berkland}\ \emph {et~al.}(1998)\citenamefont
  {Berkland}, \citenamefont {Miller}, \citenamefont {Bergquist}, \citenamefont
  {Itano},\ and\ \citenamefont {Wineland}}]{micromotion}%
  \BibitemOpen
  \bibfield  {author} {\bibinfo {author} {\bibfnamefont {D.~J.}\ \bibnamefont
  {Berkland}}, \bibinfo {author} {\bibfnamefont {J.~D.}\ \bibnamefont
  {Miller}}, \bibinfo {author} {\bibfnamefont {J.~C.}\ \bibnamefont
  {Bergquist}}, \bibinfo {author} {\bibfnamefont {W.~M.}\ \bibnamefont
  {Itano}}, \ and\ \bibinfo {author} {\bibfnamefont {D.~J.}\ \bibnamefont
  {Wineland}},\ }\href@noop {} {\bibfield  {journal} {\bibinfo  {journal} {J.
  of Appl. Phys.}\ }\textbf {\bibinfo {volume} {83}},\ \bibinfo {pages} {5025}
  (\bibinfo {year} {1998})}\BibitemShut {NoStop}%
\bibitem [{\citenamefont {Barrett}(2015)}]{MDB1}%
  \BibitemOpen
  \bibfield  {author} {\bibinfo {author} {\bibfnamefont {M.~D.}\ \bibnamefont
  {Barrett}},\ }\href@noop {} {\bibfield  {journal} {\bibinfo  {journal} {New
  Jour. Phys.}\ }\textbf {\bibinfo {volume} {17}},\ \bibinfo {pages} {053024}
  (\bibinfo {year} {2015})}\BibitemShut {NoStop}%
\bibitem [{\citenamefont {Landa}\ \emph {et~al.}(2012)\citenamefont {Landa},
  \citenamefont {Drewsen}, \citenamefont {Reznik},\ and\ \citenamefont
  {Retzker}}]{CrystalsI}%
  \BibitemOpen
  \bibfield  {author} {\bibinfo {author} {\bibfnamefont {H.}~\bibnamefont
  {Landa}}, \bibinfo {author} {\bibfnamefont {M.}~\bibnamefont {Drewsen}},
  \bibinfo {author} {\bibfnamefont {B.}~\bibnamefont {Reznik}}, \ and\ \bibinfo
  {author} {\bibfnamefont {A.}~\bibnamefont {Retzker}},\ }\href@noop {}
  {\bibfield  {journal} {\bibinfo  {journal} {New Jour. Phys.}\ }\textbf
  {\bibinfo {volume} {14}},\ \bibinfo {pages} {093023} (\bibinfo {year}
  {2012})}\BibitemShut {NoStop}%
\bibitem [{\citenamefont {Haase}(2003)}]{Haase}%
  \BibitemOpen
  \bibfield  {author} {\bibinfo {author} {\bibfnamefont {R.}~\bibnamefont
  {Haase}},\ }\href@noop {} {\bibfield  {journal} {\bibinfo  {journal} {J.
  Phys. B}\ }\textbf {\bibinfo {volume} {36}},\ \bibinfo {pages} {1011}
  (\bibinfo {year} {2003})}\BibitemShut {NoStop}%
\bibitem [{\citenamefont {Mackay}(1962)}]{Mackay}%
  \BibitemOpen
  \bibfield  {author} {\bibinfo {author} {\bibfnamefont {A.~L.}\ \bibnamefont
  {Mackay}},\ }\href@noop {} {\bibfield  {journal} {\bibinfo  {journal} {Acta.
  Crystallogr.}\ }\textbf {\bibinfo {volume} {15}},\ \bibinfo {pages} {916}
  (\bibinfo {year} {1962})}\BibitemShut {NoStop}%
\bibitem [{\citenamefont {Dzuba}()}]{Dzuba}%
  \BibitemOpen
  \bibfield  {author} {\bibinfo {author} {\bibfnamefont {V.}~\bibnamefont
  {Dzuba}},\ }\href@noop {} {}\bibinfo {howpublished} {private
  communication}\BibitemShut {NoStop}%
\bibitem [{Note1()}]{Note1}%
  \BibitemOpen
  \bibinfo {note} {Throughout we use atomic units for polarisabilities.
  Conversion to S.I. units is via the scale factor $4\pi \epsilon _0 a_0^3$
  where $a_0$ is the Bohr radius}\BibitemShut {NoStop}%
\bibitem [{\citenamefont {Kozlov}\ \emph {et~al.}(2014)\citenamefont {Kozlov},
  \citenamefont {Dzuba},\ and\ \citenamefont {Flambaum}}]{Calc1}%
  \BibitemOpen
  \bibfield  {author} {\bibinfo {author} {\bibfnamefont {A.}~\bibnamefont
  {Kozlov}}, \bibinfo {author} {\bibfnamefont {V.~A.}\ \bibnamefont {Dzuba}}, \
  and\ \bibinfo {author} {\bibfnamefont {V.~V.}\ \bibnamefont {Flambaum}},\
  }\href@noop {} {\bibfield  {journal} {\bibinfo  {journal} {Phys. Rev. A.}\
  }\textbf {\bibinfo {volume} {90}},\ \bibinfo {pages} {042505} (\bibinfo
  {year} {2014})}\BibitemShut {NoStop}%
\bibitem [{\citenamefont {Mortensen}\ \emph {et~al.}(2006)\citenamefont
  {Mortensen}, \citenamefont {Nielsen}, \citenamefont {Matthey},\ and\
  \citenamefont {Drewsen}}]{Drewsen}%
  \BibitemOpen
  \bibfield  {author} {\bibinfo {author} {\bibfnamefont {A.}~\bibnamefont
  {Mortensen}}, \bibinfo {author} {\bibfnamefont {E.}~\bibnamefont {Nielsen}},
  \bibinfo {author} {\bibfnamefont {T.}~\bibnamefont {Matthey}}, \ and\
  \bibinfo {author} {\bibfnamefont {M.}~\bibnamefont {Drewsen}},\ }\href@noop
  {} {\bibfield  {journal} {\bibinfo  {journal} {Phys. Rev. Lett.}\ }\textbf
  {\bibinfo {volume} {96}},\ \bibinfo {pages} {103001} (\bibinfo {year}
  {2006})}\BibitemShut {NoStop}%
\bibitem [{\citenamefont {Tan}\ \emph {et~al.}(1995)\citenamefont {Tan},
  \citenamefont {Bollinger}, \citenamefont {Jelenkovic},\ and\ \citenamefont
  {Wineland}}]{Bollinger1}%
  \BibitemOpen
  \bibfield  {author} {\bibinfo {author} {\bibfnamefont {J.~N.}\ \bibnamefont
  {Tan}}, \bibinfo {author} {\bibfnamefont {J.~J.}\ \bibnamefont {Bollinger}},
  \bibinfo {author} {\bibfnamefont {B.}~\bibnamefont {Jelenkovic}}, \ and\
  \bibinfo {author} {\bibfnamefont {D.~J.}\ \bibnamefont {Wineland}},\
  }\href@noop {} {\bibfield  {journal} {\bibinfo  {journal} {Phys. Rev.Lett.}\
  }\textbf {\bibinfo {volume} {75}},\ \bibinfo {pages} {4198} (\bibinfo {year}
  {1995})}\BibitemShut {NoStop}%
\bibitem [{\citenamefont {Itano}(2000)}]{ItanoQuad}%
  \BibitemOpen
  \bibfield  {author} {\bibinfo {author} {\bibfnamefont {W.~M.}\ \bibnamefont
  {Itano}},\ }\href@noop {} {\bibfield  {journal} {\bibinfo  {journal} {J. Res.
  Natl. Inst. Stand. Technol.}\ }\textbf {\bibinfo {volume} {105}},\ \bibinfo
  {pages} {829} (\bibinfo {year} {2000})}\BibitemShut {NoStop}%
\bibitem [{Note2()}]{Note2}%
  \BibitemOpen
  \bibinfo {note} {The alignment of the crystal here is by construction of the
  initial condition.}\BibitemShut {Stop}%
\bibitem [{\citenamefont {Kien}\ \emph {et~al.}(2103)\citenamefont {Kien},
  \citenamefont {Schneeweiss},\ and\ \citenamefont
  {Rauschenbeutel}}]{polarisability}%
  \BibitemOpen
  \bibfield  {author} {\bibinfo {author} {\bibfnamefont {F.~L.}\ \bibnamefont
  {Kien}}, \bibinfo {author} {\bibfnamefont {P.}~\bibnamefont {Schneeweiss}}, \
  and\ \bibinfo {author} {\bibfnamefont {A.}~\bibnamefont {Rauschenbeutel}},\
  }\href@noop {} {\bibfield  {journal} {\bibinfo  {journal} {Eur. Phys. J. D.}\
  }\textbf {\bibinfo {volume} {67}},\ \bibinfo {pages} {92} (\bibinfo {year}
  {2103})}\BibitemShut {NoStop}%
\bibitem [{Note3()}]{Note3}%
  \BibitemOpen
  \bibinfo {note} {For a single transition, if $\alpha _{2,J}$ was sufficiently
  small, it could be incorporated into the definition of $\Delta \alpha
  $}\BibitemShut {NoStop}%
\bibitem [{\citenamefont {Quinet}\ \emph {et~al.}(1999)\citenamefont {Quinet}
  \emph {et~al.}}]{lifetimes2}%
  \BibitemOpen
  \bibfield  {author} {\bibinfo {author} {\bibfnamefont {P.}~\bibnamefont
  {Quinet}} \emph {et~al.},\ }\href@noop {} {\bibfield  {journal} {\bibinfo
  {journal} {Mon. Not. R. Astron. Soc.}\ }\textbf {\bibinfo {volume} {307}},\
  \bibinfo {pages} {934} (\bibinfo {year} {1999})}\BibitemShut {NoStop}%
\bibitem [{\citenamefont {Chu}\ and\ \citenamefont {Otsuka}(2008)}]{doughnut}%
  \BibitemOpen
  \bibfield  {author} {\bibinfo {author} {\bibfnamefont {S.-C.}\ \bibnamefont
  {Chu}}\ and\ \bibinfo {author} {\bibfnamefont {K.}~\bibnamefont {Otsuka}},\
  }\href@noop {} {\bibfield  {journal} {\bibinfo  {journal} {Opt. Comm.}\
  }\textbf {\bibinfo {volume} {281}},\ \bibinfo {pages} {1647} (\bibinfo {year}
  {2008})}\BibitemShut {NoStop}%
\bibitem [{\citenamefont {Yudin}\ \emph {et~al.}(2010)\citenamefont {Yudin},
  \citenamefont {Taichenachev}, \citenamefont {Oates}, \citenamefont {Barber},
  \citenamefont {Lemke}, \citenamefont {Ludlow}, \citenamefont {Sterr},
  \citenamefont {Lisdat},\ and\ \citenamefont {Riehle}}]{hyperRamsey1}%
  \BibitemOpen
  \bibfield  {author} {\bibinfo {author} {\bibfnamefont {V.~I.}\ \bibnamefont
  {Yudin}}, \bibinfo {author} {\bibfnamefont {A.~V.}\ \bibnamefont
  {Taichenachev}}, \bibinfo {author} {\bibfnamefont {C.~W.}\ \bibnamefont
  {Oates}}, \bibinfo {author} {\bibfnamefont {Z.~W.}\ \bibnamefont {Barber}},
  \bibinfo {author} {\bibfnamefont {N.~D.}\ \bibnamefont {Lemke}}, \bibinfo
  {author} {\bibfnamefont {A.~D.}\ \bibnamefont {Ludlow}}, \bibinfo {author}
  {\bibfnamefont {U.}~\bibnamefont {Sterr}}, \bibinfo {author} {\bibfnamefont
  {C.}~\bibnamefont {Lisdat}}, \ and\ \bibinfo {author} {\bibfnamefont
  {F.}~\bibnamefont {Riehle}},\ }\href@noop {} {\bibfield  {journal} {\bibinfo
  {journal} {Phys. Rev. A.}\ }\textbf {\bibinfo {volume} {82}},\ \bibinfo
  {pages} {011804(R)} (\bibinfo {year} {2010})}\BibitemShut {NoStop}%
\bibitem [{\citenamefont {Huntemann}\ \emph
  {et~al.}(2012{\natexlab{b}})\citenamefont {Huntemann}, \citenamefont
  {Lipphardt}, \citenamefont {Okhapkin}, \citenamefont {Tamm}, \citenamefont
  {Peik}, \citenamefont {Taichenachev},\ and\ \citenamefont
  {Yudin}}]{hyperRamsey2}%
  \BibitemOpen
  \bibfield  {author} {\bibinfo {author} {\bibfnamefont {N.}~\bibnamefont
  {Huntemann}}, \bibinfo {author} {\bibfnamefont {B.}~\bibnamefont
  {Lipphardt}}, \bibinfo {author} {\bibfnamefont {M.}~\bibnamefont {Okhapkin}},
  \bibinfo {author} {\bibfnamefont {C.}~\bibnamefont {Tamm}}, \bibinfo {author}
  {\bibfnamefont {E.}~\bibnamefont {Peik}}, \bibinfo {author} {\bibfnamefont
  {A.~V.}\ \bibnamefont {Taichenachev}}, \ and\ \bibinfo {author}
  {\bibfnamefont {V.~I.}\ \bibnamefont {Yudin}},\ }\href@noop {} {\bibfield
  {journal} {\bibinfo  {journal} {Phys. Rev. Lett.}\ }\textbf {\bibinfo
  {volume} {109}},\ \bibinfo {pages} {213002} (\bibinfo {year}
  {2012}{\natexlab{b}})}\BibitemShut {NoStop}%
\bibitem [{\citenamefont {Dick}(1987)}]{Dick}%
  \BibitemOpen
  \bibfield  {author} {\bibinfo {author} {\bibfnamefont {G.~J.}\ \bibnamefont
  {Dick}},\ }\href@noop {} {\bibfield  {journal} {\bibinfo  {journal} {Proc.
  Precise Time and Time Interval}\ ,\ \bibinfo {pages} {133}} (\bibinfo {year}
  {1987})}\BibitemShut {NoStop}%
\bibitem [{\citenamefont {Itano}\ \emph {et~al.}(1993)\citenamefont {Itano},
  \citenamefont {Bergquist}, \citenamefont {Bollinger}, \citenamefont
  {Gilligan}, \citenamefont {Heinzen}, \citenamefont {Moore}, \citenamefont
  {Raizen},\ and\ \citenamefont {Wineland}}]{Itano}%
  \BibitemOpen
  \bibfield  {author} {\bibinfo {author} {\bibfnamefont {W.~M.}\ \bibnamefont
  {Itano}}, \bibinfo {author} {\bibfnamefont {J.~C.}\ \bibnamefont
  {Bergquist}}, \bibinfo {author} {\bibfnamefont {J.~J.}\ \bibnamefont
  {Bollinger}}, \bibinfo {author} {\bibfnamefont {J.~M.}\ \bibnamefont
  {Gilligan}}, \bibinfo {author} {\bibfnamefont {D.~J.}\ \bibnamefont
  {Heinzen}}, \bibinfo {author} {\bibfnamefont {F.~L.}\ \bibnamefont {Moore}},
  \bibinfo {author} {\bibfnamefont {M.~G.}\ \bibnamefont {Raizen}}, \ and\
  \bibinfo {author} {\bibfnamefont {D.~J.}\ \bibnamefont {Wineland}},\
  }\href@noop {} {\bibfield  {journal} {\bibinfo  {journal} {Phys. Rev. A.}\
  }\textbf {\bibinfo {volume} {47}},\ \bibinfo {pages} {3554} (\bibinfo {year}
  {1993})}\BibitemShut {NoStop}%
\bibitem [{\citenamefont {Dube}\ \emph {et~al.}(2014)\citenamefont {Dube},
  \citenamefont {Madej}, \citenamefont {Tibbo},\ and\ \citenamefont
  {Bernard}}]{MagicSr}%
  \BibitemOpen
  \bibfield  {author} {\bibinfo {author} {\bibfnamefont {P.}~\bibnamefont
  {Dube}}, \bibinfo {author} {\bibfnamefont {A.~A.}\ \bibnamefont {Madej}},
  \bibinfo {author} {\bibfnamefont {M.}~\bibnamefont {Tibbo}}, \ and\ \bibinfo
  {author} {\bibfnamefont {J.~E.}\ \bibnamefont {Bernard}},\ }\href@noop {}
  {\bibfield  {journal} {\bibinfo  {journal} {Phys. Rev. Lett.}\ }\textbf
  {\bibinfo {volume} {112}},\ \bibinfo {pages} {173002} (\bibinfo {year}
  {2014})}\BibitemShut {NoStop}%
\bibitem [{\citenamefont {Bollinger}\ \emph {et~al.}(1994)\citenamefont
  {Bollinger}, \citenamefont {Wineland},\ and\ \citenamefont
  {Dubin}}]{Bollinger2}%
  \BibitemOpen
  \bibfield  {author} {\bibinfo {author} {\bibfnamefont {J.}~\bibnamefont
  {Bollinger}}, \bibinfo {author} {\bibfnamefont {D.~J.}\ \bibnamefont
  {Wineland}}, \ and\ \bibinfo {author} {\bibfnamefont {D.~H.~E.}\ \bibnamefont
  {Dubin}},\ }\href@noop {} {\bibfield  {journal} {\bibinfo  {journal} {Phys.
  Plasmas}\ }\textbf {\bibinfo {volume} {1}},\ \bibinfo {pages} {1403}
  (\bibinfo {year} {1994})}\BibitemShut {NoStop}%
\bibitem [{\citenamefont {Safranova}\ \emph {et~al.}(2012)\citenamefont
  {Safranova}, \citenamefont {Kozlov},\ and\ \citenamefont
  {Clark}}]{Safranova}%
  \BibitemOpen
  \bibfield  {author} {\bibinfo {author} {\bibfnamefont {M.~S.}\ \bibnamefont
  {Safranova}}, \bibinfo {author} {\bibfnamefont {M.~G.}\ \bibnamefont
  {Kozlov}}, \ and\ \bibinfo {author} {\bibfnamefont {C.~W.}\ \bibnamefont
  {Clark}},\ }\href@noop {} {\bibfield  {journal} {\bibinfo  {journal} {IEEE
  Trans. on Ultr. Ferro. and Freq. Con.}\ }\textbf {\bibinfo {volume} {59}},\
  \bibinfo {pages} {439} (\bibinfo {year} {2012})}\BibitemShut {NoStop}%
\bibitem [{\citenamefont {Sahoo}\ \emph {et~al.}(2009)\citenamefont {Sahoo},
  \citenamefont {Timmermans}, \citenamefont {Das},\ and\ \citenamefont
  {Mukherjee}}]{Sahoo}%
  \BibitemOpen
  \bibfield  {author} {\bibinfo {author} {\bibfnamefont {B.~K.}\ \bibnamefont
  {Sahoo}}, \bibinfo {author} {\bibfnamefont {R.~G.~E.}\ \bibnamefont
  {Timmermans}}, \bibinfo {author} {\bibfnamefont {B.~P.}\ \bibnamefont {Das}},
  \ and\ \bibinfo {author} {\bibfnamefont {D.}~\bibnamefont {Mukherjee}},\
  }\href@noop {} {\bibfield  {journal} {\bibinfo  {journal} {Phys. Rev. A.}\
  }\textbf {\bibinfo {volume} {80}},\ \bibinfo {pages} {062506} (\bibinfo
  {year} {2009})}\BibitemShut {NoStop}%
\end{thebibliography}%
\end{document}